\documentclass[twocolumn,floatfix,aps,amsmath,amssymb,amsfonts,10pt,showpacs,prd]{revtex4-1}
\usepackage[dvips]{graphicx}
\usepackage{epsfig}
\begin{document}
\title{Light front Casimir effect at finite temperature}

\author{P. L. M. Rodrigues}
\email{pennlee@ufpa.br}
\author{Silvana Perez}
\email{silperez@ufpa.br}
\author{Danilo T. Alves}
\email{danilo@ufpa.br}
\author{Van S\'{e}rgio Alves}
\email{vansergi@ufpa.br}
\affiliation{Faculdade de F\'{\i}sica, Universidade Federal do Par\'a, 66075-110, Bel\'{e}m, Par\'a, Brazil}
\author{Charles R. Silva}
\email{charles.rocha@ifpa.edu.br}
\affiliation{Instituto Federal do Par\'{a}, 66093-020, Bel\'{e}m, Par\'a, Brazil}
\date{\today}

\begin{abstract} 
The correct description of the standard Casimir effect for periodic boundary conditions via light front formalism implies in these conditions imposed at fixed Minkowski times [Almeida {\it et al.} Phys. Rev. {\bf D 87}, 065028 (2013); Chabysheva and Hiller, Phys. Rev. {\bf D 88}, 085006 (2013)] instead of fixed light front times. The unphysical nature of this latter condition is manifested in the vacuum part by no regularization yielding a finite Casimir energy density [Lenz and Steinbacher, Phys. Rev. {\bf D 67}, 045010 (2003)]. In the present paper, we extend this discussion and analyze the problem of the light front quantization with simultaneous presence of a thermal bath and boundary conditions. Considering both the oblique light front as well as Dirac light front coordinates, we show that the imposition of periodic boundary conditions at fixed Minkowski times recovers the expected behaviors for the energy density and Casimir entropy. We also investigate how the unphysical nature of the periodic boundary conditions imposed at fixed light front times manifests in the thermal part of the energy and entropy, showing that 
in the classical limit the Casimir entropy decreases linearly with the temperature (not becoming independent of the temperature as expected), and also that the Kirchhoff theorem is not respected.
\end{abstract}

\pacs{03.70.+k, 11.10.Wx}
\maketitle

\section{Introduction}
The study of the Casimir effect in light front dynamics has been investigated by several authors \cite{lenz,casimir-lf,hiller}.
In the context of the non-massive scalar field in $(3+1)$ dimensions and imposing a periodic boundary condition on
the longitudinal light front coordinate, and at fixed light front times, the traditional methods of regularization fail in extracting the finite part of the vacuum energy density \cite{lenz}. 
These problematic situations can be avoided if, even for the quantization taken
on equal light front times, the periodic boundary condition is taken on equal Minkowski times \cite{casimir-lf}. This prescription of boundary
conditions on equal usual times consistently leads, in the light front formalism, to the usual Casimir energy density found in literature \cite{casimir-lf,hiller}.

The problem of light front quantization with simultaneous presence of a thermal bath and
boundary conditions was investigated in Refs. \cite{lenz,casimir-lf}. 
Specifically, the authors in \cite{casimir-lf} considered oblique light front coordinates and periodic boundary condition on
the longitudinal light front coordinate, and at fixed light front times, and obtained 
that, beyond the absence of regularization and break of the isotropy,
the imposition of periodic boundary conditions on 
$x^0+x^3=const.$ hyperplanes breaks the temperature inversion symmetry.

In the present paper, we extend the investigation started in Ref. \cite{casimir-lf} on the problem of 
the light front quantization with simultaneous presence of a thermal bath and
boundary conditions. We consider both usual Dirac light front coordinates as well as oblique light front coordinates and obtain exact formulas for the energy density and entropy for periodic boundary condition at fixed Minkowski times. 
We also investigate how the unphysical nature of the periodic boundary conditions imposed at fixed light front times manifests in the thermal part of the energy and entropy. 

The paper is organized as follows. In Sec. \ref{oblique-coordinates}, we consider the oblique light front coordinates and calculate the energy density and entropy of the system, subject to boundary conditions imposed at fixed light front times and fixed Minkowski times and discuss the TIS.  The classical limit is also considered. In Sec.  \ref{dirac-coordinates} we discuss the thermal effects when the Dirac light front coordinates are considered. The final remarks are given in Sec. \ref{conclusions}. Throughout the paper we consider the natural units where $\hbar=c=k_B=1$.


\section{Oblique Light Front Coordinates}
\label{oblique-coordinates}

In the context of the instant form formalism, the Casimir entropy and the Helmholtz free energy have been discussed in many papers \cite{casimir-entropy-equal-times, plunien, sit, feinberg, revzen, rubin, silva}.  
In this Section, considering the  light front dynamics, we investigate these quantities, under boundary conditions imposed at fixed Minkowski times and light front times. 
The inclusion  of thermal effects in light front (LF) dynamics as well as applications of this formalism have been investigated in many publications, including Refs. \cite{elser, brodsky,elser,beyer, alves, weldon-1, weldon-2, das-zhou, das-perez-glf, proestos, strauss, strauss-mattiello, das-perez-unruh, perez-rocha-polar-tensor}. 



As discussed in \cite{weldon-1, weldon-2, casimir-lf}, a convenient way to include thermal effects in light front dynamics is using the oblique light front coordinates. This new system of coordinates $\bar{x}$ is related to the  Minkowski coordinates $x$  through a linear transformation
\begin{equation}\label{eq02}
\bar{x}^{\mu} = {L^{\mu}}_{\nu} x^{\nu},
\end{equation}
where
\begin{equation}\label{eq02-II}
{L^{\mu}}_{\nu} = \begin{pmatrix} 1 & 0& 0& 1\\
0 & 1& 0& 0 \\
0& 0& 1& 0 \\
0 & 0& 0& 1
\end{pmatrix}.
\end{equation}

We then consider the massless scalar field in the oblique LF coordinate system, given by the lagrangian density
\begin{eqnarray}
{\cal L} &=& - \bar{\partial}_{0} \phi \bar{\partial}_{3} \phi - \frac{1}{2} (\bar{\partial}_{\alpha} \phi)^2 - \frac{1}{2}  (\bar{\partial}_{3} \phi)^2, \qquad \alpha=1,2.  \nonumber
\end{eqnarray}
The field decomposition takes the form \cite{das-perez-glf}
\begin{eqnarray}
\phi(\bar{x})&=&\frac{1}{(2 \pi)^{3/2}} \int_{-\infty}^{+\infty} d^2\bar{k}_{\alpha} \nonumber \\
&\times& \int_0^{\infty} \frac{d \bar{k}_3}{2 \bar{k}_3} \left[e^{- i \tilde{\bar{k}} \cdot \bar{x}}a(\bar{k}) + e^{+ i \tilde{\bar{k}} \cdot \bar{x}}a^{\dagger}(\bar{k})\right],
\end{eqnarray}
with $\tilde{\bar{k}}\equiv(\bar{k}_0, - \bar{k}_{\alpha}, - \bar{k}_3)$ and
\begin{equation}
\bar{k}_0\equiv\frac{\bar{k}_{\alpha}^2+ \bar{k}_3^2}{2 \bar{k}_3} > 0.
\end{equation}
Finally, $a^{\dagger}$ and  $a$ are the creation and annihilation operators, respectively, satisfying
\begin{equation}
[a(\bar{k}),a^{\dagger} (\bar{q}) ] = 2 \bar{k}_3 \delta^3(\bar{k}- \bar{q}).
\end{equation}

Next, we calculate  the Helmholtz free energy and Casimir entropy, as functions of the temperature, for a periodic boundary condition taken at fixed Minkowski times,
\begin{equation}
\phi(\bar{x}^0, \bar{x}^{\alpha}, \bar{x}^3)=\phi(\bar{x}^0+L, \bar{x}^{\alpha}, \bar{x}^3+L).
\label{fixed-equal-times-bc}
\end{equation}

We then follow \cite{casimir-lf} (and references therein) and write the Helmholtz free energy $F$ as
\begin{eqnarray}\label{free-ener}
F&\equiv& F_0 + F_{\beta},
\end{eqnarray}
where $F_{0}$ is the vacuum energy density, given by
\begin{equation}\label{free-ener-0}
F_{0} = \langle 0 | H |0 \rangle\,
\end{equation}
($H$ is the hamiltonian), and
$F_{\beta}$ is the explicit finite temperature contribution,
\begin{equation}\label{free-ener-beta}
F_{\beta} = T \sum_{n_k} \ln (1 - e^{-{\beta} \bar{k}_0}).
\end{equation}
Furthermore, $\bar{k}_0$ is the energy of the statistical system.
Considering first the boundary condition (\ref{fixed-equal-times-bc}), we find the following constraint for the momenta:
\begin{equation}\label{constraint}
\frac{\bar{k}_{\alpha}^2}{2 \bar{k_3}} - \frac{\bar{k}_3}{2} = \frac{2 \pi n }{L}, \qquad n= 0, \pm  1, \pm 2, \dots
\end{equation}
and applying it in the Helmholtz free energy we then write
\begin{eqnarray}
F_{\beta} &=& \frac{T \cal{A}}{(2 \pi)^2} \sum_{n=-\infty}^{+ \infty} \int_{-\infty}^{+ \infty} d^2 \bar{k}_{\alpha} \nonumber \\
&\times& \int_0^{\infty} d  \bar{k}_3\ln (1 - e^{- \beta \bar{k}_0}) \delta \left( \frac{\bar{k}_{\alpha}^2}{2 \bar{k_3}} - \frac{\bar{k}_3}{2}  - \frac{2 \pi n}{L}\right), \nonumber \\
\end{eqnarray}
where $\cal{A}$ stands for the area associated with the transversal directions. Now, considering that the distribution  function $n_B$ acts as a natural regulator making all thermal results ultraviolet finite, we can change the variables
\begin{eqnarray}\label{var-change}
\bar{k}_3 &=& k_3 + E_k, \qquad E_k = \sqrt{k_{\alpha}^2 + k_3^2}, \nonumber \\
\bar{k}_{\alpha} &=& k_{\alpha},
\end{eqnarray}
such that the range of $k_3$ is $-\infty <k_3<\infty$, and rewrite the free Helmholtz energy as:
\begin{eqnarray}\label{eq10}
F_{\beta} &=& \frac{T \cal{A}}{(2 \pi)^2} \sum_{n=-\infty}^{+ \infty} \int_{-\infty}^{+\infty} d^3 {k} \ln (1 - e^{- \beta E_k}) \delta \left(k_3 + \frac{2 \pi n}{L}\right) \nonumber \\
&=& \frac{ T \cal{A}}{(2 \pi)^2} \sum_{n=-\infty}^{+ \infty} \int_{-\infty}^{+\infty} d^2 {k_{\alpha}} \ln (1 - e^{- \beta E_k^n}),
\end{eqnarray}
where in the intermediate steps we integrated over $k_3$. We also defined the discretized energy $E_k^n$ as
\begin{equation}
E_k^n\equiv \sqrt{k_{\alpha}^2 + \left(\frac{2 \pi n}{L}\right)^2}.
\end{equation}
The calculation now is straightforward and, without going into technical details, one can find that the finite temperature contribution to the Helmholtz free energy, when the boundary condition (\ref{fixed-equal-times-bc}) is considered,  is given by
\begin{equation}\label{free-ener-correc}
F_{\beta} = - \frac{V \zeta(4)}{\pi^2}\frac{1}{\beta^4} - \frac{2 \cal{A}}{\pi^2 L^3} \sum_{\nu, k=1}^{\infty} \frac{\xi^4}{(\nu^2+ (k \xi)^2)^2},
\end{equation}
where $V = \mathcal{A} L$ is the volume enclosed by the boundary condition, $\zeta$ stands for the Riemann zeta function  and $\xi =LT$ is a dimensionless variable. The first term is the free field configuration and after subtracting it we are left with the Casimir free energy, given by
\begin{equation}\label{correct-free-energy}
F_{\beta, \mbox{c}} = - \frac{2 \cal{A}}{\pi^2 L^3} \sum_{\nu, k=1}^{\infty} \frac{\xi^4}{(\nu^2+ (k \xi)^2)^2}.
\end{equation}
Before calculating the entropy of the system, it is worth noting that $F_{\beta, \mbox{c}}$, given by Eq.(\ref{correct-free-energy}) satisfies the TIS, expressed as \cite{sit}%
\begin{equation}
f( \xi) = \xi^4 f( 1/\xi),
\end{equation}
where $f( \xi)$ is dimensionless function defined as
\begin{equation}
f( \xi) = \frac{L^3 F_{\beta, \mbox{c}}}{\cal{A}}.
\end{equation}

This result has to be contrasted with that found for the Casimir free energy at fixed light front times,
\begin{equation}
\phi(\bar{x}^0, \bar{x}^{\alpha}, \bar{x}^3)=\phi(\bar{x}^0, \bar{x}^{\alpha}, \bar{x}^3+L),
\label{fixed-lf-times-bc}
\end{equation}
given by
\begin{equation}\label{eq.31}
F_{\beta, \mbox{c}} = - \frac{4 \mathcal{A}}{\pi^2 L^3} \sum_{\nu, k=1}^{\infty} \frac{\xi^4(\nu^2-(2\xi k)^2)}{\nu^2(\nu^2+ (2\xi k)^2)^2},
\end{equation}
which does not respect TIS (in fact, this behavior was already pointed out in Ref.\cite{casimir-lf}, where the high and low temperature limits of the free energy where evaluated).

To obtain the  Casimir entropy, we use the  definition:
\begin{equation}\label{entrop}
S_{\mbox{c}}\equiv-\frac{\partial F_{\beta, \mbox{c}}}{\partial T}.
\end{equation}

Thus, the Casimir entropy for the boundary condition (\ref{fixed-equal-times-bc}) is calculated using  the free energy (\ref{correct-free-energy}) and is written as
\begin{equation}\label{eq.30}
S_{\mbox{c}} =  \frac{8 \mathcal{A}}{\pi^2 L^2} \sum_{k, \nu = 1}^{\infty} \frac{\nu^2 \xi^3}{(\nu^2 + k^2\xi^2)^3}.
\end{equation}
On the other hand, to obtain the Casimir entropy for the boundary condition  (\ref{fixed-lf-times-bc}), we use the free energy given in Eq.(\ref{eq.31}) and find
\begin{equation}\label{eq.40}
S_{\mbox{c}} = \frac{4 \cal{A}}{\pi^2 L^2}\frac{\partial}{\partial \xi} \left(\xi^4 \sum_{k,\nu=1}^{\infty}\frac{\nu^2 - (2 \xi k)^2}{ [\nu^2 + (2 \xi k)^2]^2}\right).
\end{equation}

\begin{figure}[htb]
\centering                  
\includegraphics[scale=0.7]{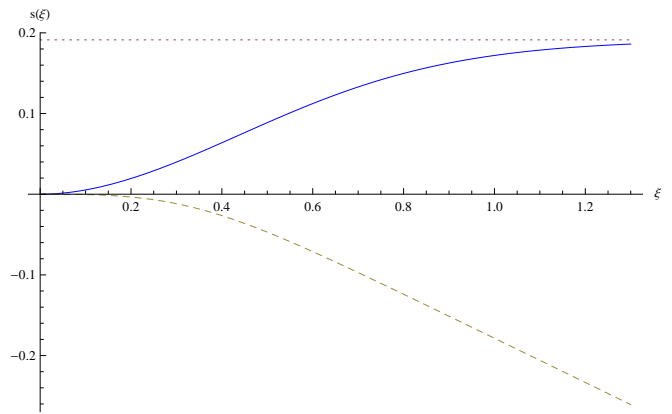}                                         
\caption{
The behavior of the dimensionless Casimir entropy $s(\xi)=L^2S(\xi)/{\cal A}$  as a function of $\xi$ .
The full line corresponds to boundary condition (\ref{fixed-equal-times-bc}); the dashed line to boundary condition (\ref{fixed-lf-times-bc}); and the dotted line to $\zeta(3)/(2\pi)$.}\label{graph-entr}  
\end{figure}

The behavior of the Casimir entropy as a function of $\xi$ is shown in Fig.(\ref{graph-entr}), for Eqs.(\ref{eq.30}) and (\ref{eq.40}).
In particular, in the classical limit  (i.e. for high temperatures,  wherein the energy equipartition theorem is valid \cite{rubin}),  $LT\gg 1$, the Casimir entropy  at fixed light front times $\bar{x}^0$, Eq.(\ref{eq.40}), decreases linearly with the temperature,
\begin{eqnarray}\label{entropy-bad}
\frac{S_{\mbox{c}}}{{\cal A} L}&=&\frac{\zeta(3)}{4\pi L^3} - \frac{\pi^2}{36 L^2}T,
\end{eqnarray}
not having the expected thermodynamical behavior in the classical limit \cite{revzen}. This has to be contrasted with  the Eq.(\ref{eq.30}), considered at fixed Minkowski times, where the entropy behaves in the expected way, increasing monotonically as the temperature increases, up to a critical value. After this value (high temperature limit), it becomes independent of the temperature,
\begin{equation}\label{entropy-good}
\frac{S_{\mbox{c}}}{{\cal A} L}=\frac{\zeta(3)}{2\pi L^3},
\end{equation}
being exactly the known classical entropy per volume unit \cite{casimir-entropy-equal-times}.
%
 

In addition to the analysis of the entropy, we also examine the Casimir energy in the classical limit for the two boundary conditions given by (\ref{fixed-lf-times-bc}) and (\ref{fixed-equal-times-bc}).

The Casimir energy at finite temperature is defined as a sum of the zero-point energy and the Casimir energy of the quantum system
referring to real occupied states:
\begin{equation}
E_{\mbox{c}}(L)=E_{\mbox{c},0}(L)+{E}_{\mbox{c},\beta}(L),
\end{equation}
where $E_{\mbox{c},0}(L)$ is the regularized vacuum energy density and ${E}_{\mbox{c},\beta}(L)$ is the explicit temperature-dependent contribution to the Casimir energy density, which is obtained from difference between the energy density of the field in the presence of external constraints and the one of the free field, i.e.,
\begin{equation}\label{cor-ener}
E_{\mbox{c},\beta}(L)=E_{\beta}(L)-E_{\beta}(\infty).
\end{equation}
Furthermore, in the Ref.\cite{plunien} the authors noted that ${E}_{\mbox{c},\beta}(L)$, in the high temperature limit, contains exactly the zero-temperature Casimir energy with opposite sign,
\begin{equation}
{E}_{\mbox{c},\beta}(L)=-E_{\mbox{c},0}(L)+\mathcal{O}\left(e^{-LT}\right),
\end{equation}
leading to a vanishing total Casimir energy,
\begin{equation}
E_{\mbox{c}}(L)= 0,
\end{equation}
where higher order terms vanish exponentially fast. This result is also known as the Kirchhoff theorem \cite{feinberg} and can also be understood as been a consequence of the one-to-one correspondence \cite{hushwater} together with the classical energy equipartition theorem.

Now, if we consider the boundary condition (\ref{fixed-equal-times-bc}) the energy ${E}_{\beta}(L)$ can be calculated from $F_{\beta}$ through 
\begin{equation}\label{evaluate-ene}
E_{\beta}(L) = - \frac{\partial}{\partial \beta} \left(\beta F_{\beta}\right),
\end{equation}
and matches exactly the instant form calculation \cite{plunien}. In particular, in the high temperature limit is given by
\begin{equation}\label{correct}
E_{\beta}(L)= \frac{{\cal A}}{\pi^2 L^3}\zeta(4)+\frac{3V}{\pi^2}T^4\zeta(4).
\end{equation}
It is worth emphasizing that the temperature independent term appearing in this expression is exactly the zero-temperature Casimir energy with opposite sign \cite{casimir-lf}.

The free field configuration is obtained replacing the summation in Eq.(\ref{free-ener-beta}) by an integration over all momenta $k$ \cite{plunien}. Therefore, one obtains the energy of the free field configuration as 
\begin{eqnarray}
{E}_{\beta}(\infty) &=& \frac{3 V }{\pi^2}T^4\zeta(4).
\end{eqnarray}
Collecting all the results we find 
\begin{equation}\label{kirchhoff}
{E}_{\mbox c}(L) = 0,
\end{equation}
respecting the Kirchhoff theorem.

On the other hand, if we consider the boundary condition (\ref{fixed-lf-times-bc}) and follow the same  steps,  the high temperature limit of ${E}_{\beta}(L)$ is obtained to be
\begin{equation}
{E}_{\beta}(L)= \frac{3 \cal{A}}{16 \pi^2 L^3} \zeta(4) + \frac{3 V T^4 \zeta(4)}{\pi^2} - \frac{{\cal A} T^2}{2 \pi^2 L} (\zeta(2))^2.
\end{equation}
Assuming that this expression contains exactly the zero-temperature Casimir energy with opposite sign, we identify the temperature-independent term with the Casimir energy at zero temperature, namely
\begin{equation}
E_{\mbox{c},0}(L)= -\frac{3 \cal{A}}{16 \pi^2 L^3} \zeta(4).
\end{equation}

The free field configuration is given by
\begin{equation}
{E}_{\beta}(\infty) = \frac{V}{ (2 \pi)^3} \int d^2 k \int_0^{\infty} \frac{dk_3}{2 k_3}\frac{k_{\alpha}^2 + k_3^2}{e^{\beta \frac{k_{\alpha}^2 + k_3^2}{2 k_3}}-1}.
\end{equation}
A straightforward calculation gives the result
\begin{equation}
E_{\beta}(\infty)= \frac{3 V T^4 \zeta(4)}{\pi^2}.
\end{equation}
Collecting all the results we find 
\begin{equation}\label{kir-bad}
E_{\mbox{c}}(L)= - \frac{{\cal A} T^2}{2 \pi^2 L} (\zeta(2))^2,
\end{equation}
which clearly does not respect the Kirchhoff theorem.


\section{Dirac light front coordinates}
\label{dirac-coordinates}

For completeness, in  this section we present the analysis of the problem considering the Dirac light front coordinates. These coordinates were first proposed by Dirac, and are given by
\begin{equation}\label{dirac-coords}
\bar{x} = (x^+, x^{\alpha}, x^-),
\end{equation}
where $x^{\alpha}$ are the Minkowski coordinates and we have used the notation $x^{\pm} = x^0 \pm x^3$. The canonical quantization of the Klein-Gordon field is straightforward (see for example \cite{das-perez-glf}), such that the field decomposition takes the form
\begin{eqnarray}\label{field-decomp-dirac}
\phi(\bar{x})&=&\frac{1}{(2 \pi)^{3/2}} \int_{-\infty}^{+\infty} d^2 k_{\alpha} \nonumber \\
&\times& \int_0^{\infty} \frac{d {k}_-}{2 {k}_-} \left[e^{- i \bar{k} \cdot \bar{x}}a(\bar{k}) + e^{+ i \bar{k} \cdot \bar{x}}a^{\dagger}(\bar{k})\right],
\end{eqnarray}
with $\bar{k}\equiv(k_+, k_{\alpha}, k_-)$ and
\begin{equation}
k_+\equiv\frac{k_{\alpha}^2}{4 k_-} > 0.
\end{equation}
Finally, $a^{\dagger}$ and  $a$ are the creation and annihilation operators, respectively, satisfying
\begin{equation}
[a(\bar{k}),a^{\dagger} (\bar{q}) ] = 2 {k}_- \delta^3(\bar{k}- \bar{q}).
\end{equation}
Next, we calculate  the Helmholtz free energy and Casimir entropy, as functions of the temperature, for a periodic boundary condition taken at fixed Minkowski times,
\begin{equation}
\phi(x^+ + L, x^{\alpha}, x^- - L)=\phi(x^+ , x^{\alpha}, x^- ).
\label{fixed-equal-times-bc-dirac}
\end{equation}
Applying this boundary condition in the field (\ref{field-decomp-dirac}) gives the following constraint for the momenta
\begin{equation}
k_+ - k_- = \frac{2 \pi n}{L}, \qquad n=0,\pm 1, \pm 2, \cdots
\end{equation}

To include thermal effects in the analysis, we  use the manifestly covariant description of thermal field theories \cite{israel, weldon-3}. In this case,  the ensemble average has the form
\begin{equation}\label{eq.01}
\left\langle \mathcal{O}\right\rangle_\beta = \mbox{Tr}\rho \mathcal{O},
\end{equation}
where 
\begin{equation}\label{rho}
\rho = e^{-\beta u\cdot k} = e^{-\beta u^{\mu} k_{\mu}},
\end{equation}
is density matrix and we are considering a Lorentz frame in which the heatbath has the velocity $u^{\mu}$, normalized to unit. In conventional thermal field theory, where the metric is diagonal and is of the form $(+,-,-,\cdots, -)$, one can choose a rest frame of the heat bath corresponding to $u^{\mu} = (1, 0,0, \cdots, 0)$ and in this case, Eq.(\ref{eq.01}) would reduce to the conventional definition of ensemble average used in instant form calculations. In contrast, once a light-front description of a theory is manifestly relativistic, we consider an alternative velocity for the heat bath, given by
\begin{equation}
u^\mu =(1,0,0,1),
\end{equation}
in which case, the density matrix is written as
\begin{equation}
\rho =  e^{-\frac{\beta }{2}(k_+ + k_-)}.
\end{equation}
Using this prescription, the explicit finite temperature contribution to the  Helmholtz free energy becomes
\begin{eqnarray}
F&=& - \frac{1}{\beta}\mbox{ln}\mbox{Tr}\rho\nonumber\\
&=& \frac{2TA}{(2\pi)^2}\displaystyle\sum_{n=-\infty}^{+\infty}\int d^2k_{\alpha} \nonumber \\
&\times & \int_{0}^{\infty}dk_{-}\mbox{ln}\left[1-e^{-\beta(k_+ + k_-)}\right] \delta\left(\frac{k_i^2}{4k_{-}} - \frac{2\pi n}{L}\right).\nonumber \\
\end{eqnarray}
We then change the variables,
\begin{eqnarray*}
k_- &=& \frac{1}{2} (k_3 + E_k), \qquad E_k=\sqrt{k_3^2 + k_i^2}.
\end{eqnarray*}
such that the range of $k_3$  is $-\infty <k_3 <\infty$ and rewrite the free Helmholtz energy as:
\begin{eqnarray}\label{eq10-II}
F_\beta & = & \frac{TA}{(2\pi)^2}\displaystyle\sum_{n=-\infty}^{+\infty}\int d^2k_{\alpha}\int_{0}^{\infty}dk_{3}\mbox{ln}\left(1-e^{-\beta E_k}\right)\nonumber\\
& & \times \delta\left(k_3 + \frac{2\pi n}{L}\right) \nonumber \\ 
 &=& \frac{TA}{(2\pi)^2}\displaystyle\sum_{n=-\infty}^{+\infty}\int d^2k_{\alpha} \mbox{ln}\left(1-e^{-\beta E_k^n}\right),
\end{eqnarray}
recovering Eq.(\ref{eq10}), which reproduces the correct behavior for the free energy and entropy. 

%
%


\section{Final Remarks}
\label{conclusions}

In the present paper, we investigated the problem of the light front quantization with simultaneous presence of a thermal bath and periodic boundary conditions, extending previous investigations found in the literature \cite{lenz, casimir-lf, hiller}. Considering both the oblique light front (Eqs. (\ref{eq02}) and (\ref{eq02-II})) as well as Dirac light front coordinates (Eq. (\ref{dirac-coords})), we showed that the imposition of periodic boundary conditions at fixed Minkowski times (Eqs. (\ref{fixed-equal-times-bc}) and (\ref{fixed-equal-times-bc-dirac})) leads to Eq. (\ref{eq10}) (or Eq. (\ref{eq10-II})), from which one can recover the expected behaviors for the energy density and Casimir entropy.

Since the manner by which the unphysical nature of periodic boundary conditions imposed at fixed light front times manifests in the
vacuum fluctuations is well described in the literature \cite{lenz}, here we also investigated how this unphysical nature manifests in the thermal part of the energy and entropy. We showed that in the classical limit the Casimir entropy decreases linearly with the temperature (Eq. (\ref{entropy-bad})), not becoming independent of the temperature as expected (Eq. (\ref{entropy-good})). Moreover, we showed that the Kirchhoff theorem is not respected (Eq. (\ref{kir-bad})).

\acknowledgments

The authors acknowledge the Referees for comments and suggestions.
This work was supported by CNPq and CAPES (Brazil).

\end{document}